\documentclass[conference]{IEEEtran}
\IEEEoverridecommandlockouts

\usepackage{cite}
\usepackage{amsmath,amssymb,amsfonts}
\usepackage{graphicx}
\usepackage{textcomp}
\usepackage{xcolor}
\usepackage{verbatim}
\usepackage{fancybox}
\usepackage{tikz}
\usepackage{amsmath}
\usepackage{algorithm}
\usepackage{wrapfig}
\usepackage[noend]{algpseudocode}

\newcommand{\CloudBuild}{{\sc{CloudBuild}}}
\newcommand{\ie}{\textit{i.e.,}}
\newcommand{\eg}{\textit{e.g.,}}
\newcommand{\etc}{\textit{etc.}}

\newcommand{\figref}[1]{Figure~\ref{#1}}

\newcommand{\algoref}[1]{Algorithm~\ref{#1}}

\begin{document}

\title{Towards Predicting the Impact of Software Changes on Building Activities}

\author{\IEEEauthorblockN{Michele Tufano}
\IEEEauthorblockA{\textit{The College of William and Mary}\\
Williamsburg, Virginia, USA \\
mtufano@cs.wm.edu}
\and
\IEEEauthorblockN{Hitesh Sajnani}
\IEEEauthorblockA{\textit{Microsoft}\\
Redmond, Washington, USA \\
hitsaj@microsoft.com}
\and
\IEEEauthorblockN{Kim Herzig}
\IEEEauthorblockA{\textit{Microsoft}\\
Redmond, Washington, USA \\
kimh@microsoft.com}
}

\maketitle

\begin{abstract}
The pervasive adoption of Continuous Integration practices -- both in industry and open source projects -- has led software building to become a daily activity for thousands of developers around the world. Companies such as Microsoft have invested in in-house infrastructures with the goal of optimizing the build process. \CloudBuild, a distributed and caching build service developed internally by Microsoft, runs the build process in parallel in the cloud and relies on caching to accelerate builds. This allows for agile development and rapid delivery of software even several times a day. However, moving towards faster builds requires not only improvements on the infrastructure side, but also attention to developers' changes in the software. Surely, architectural decisions and software changes, such as addition of dependencies, can lead to significant build time increase. Yet, estimating the impact of such changes on build time can be challenging when dealing with complex, distributed, and cached build systems. 

In this paper, we envision a predictive model able to preemptively alert developers on the extent to which their software changes may impact future building activities. In particular, we describe an approach that analyzes the developer's change and predicts (i) whether it impacts (any of) the Longest Critical Path; (ii) may lead to build time increase and its delta; and (iii) the percentage of future builds that might be affected by such change.
\end{abstract}


\section{Introduction}
\label{sec:intro}
Compiling and verifying software is a process involving several types of \textit{tasks} with the end goal of translating source code into an efficient executable program. The process usually starts by fetching source code from a repository managed by a version control system such as Git, Subversion, Bazaar, \etc.
Next, optional static program analysis tasks can be performed in order to assess code quality requirements defined in the organization's build process. These include, but not limited to, security vulnerability checks, adherence of the code to stylistic and formatting rules, comments, code clones, and code quality metrics checks such as cohesion and complexity.
Subsequently, source code is compiled and turned in executable (or intermediate) objects that are combined to generate potentially different versions of the executable program. Automated unit test cases execute in parallel to compilation tasks to ensure code quality. Finally, additional tasks may be executed such as storage of the program drops, cleaning of temporary files, logs, and notifications.

Continuous Integration (CI) is a development practice that involves frequent integration of code changes into a shared repository. Developers are encouraged to integrate often and daily, while each integration is verified by an automated build and tests \cite{fowler2006continuous}. CI aims to avoid the problems caused by a separate integration phase in the software process: unpredictability and large integration effort \cite{fowler2006continuous, 7284593}. The adoption of CI has been steadily growing, both in industry \cite{feitelson2013development, CLAPS201521} and open source projects \cite{holck2003continuous}, thanks to its ability to facilitate agile software development and allowing faster delivery cadence of software products. On the other hand, building software daily, several times a day, potentially for many different developers and teams, requires fast and reliable builds.

Companies such as Microsoft, Google, and Facebook have invested in infrastructures with the goal of accelerating the build and verification processes to enable their teams to build, integrate, and iterate faster---a procedure to ensure teams and products stay competitive. All of these modern build system rely on two main principles: distribution and pluralization as well as caching. Distribution build tasks minimizes resource limitation while caching reduces the amount of resource to be spent in the first place. All of these modern build systems helped teams to accelerate their development process. In fact, nowadays, build times no longer depend on infrastructure and resources availability, but on architectural constraints. The more dependencies a software project has between its individual components, the lower the ability to make use of parallelism (we can only compile two independent components in parallel) and caching (build targets depending on recompiled tasks cannot come from cache). We provide more details on these concepts later in this paper.

In other words, optimizing build speed nowadays becomes more and more a question of designing software systems and dependency structures to allow modern build systems to make full use of distribution and caching. Every code change adding a dependency between two previously independent software modules can impact build speed and slow down a development teams release cycles.  

In this paper, we shifted the focus of build performances on the \textit{developer-side}, by envisioning an approach able to alert developers on the extent to which their software changes may impact future building activities. The goal is to empower developers by raising awareness of the impact of their software changes. Developers can then decide whether to perform corrective operations or confirm the current change. Such approach could be integrated in the Pull Request (PR) process, where code changes are reviewed not only using classic guidelines, but also on the impact of these changes to build time \cite{wen2018icsme}. Changes that are likely to have a significant impact on build speed may need special approval and may trigger more carefully code reviews whether the newly introduced dependencies are actual necessary or whether solutions with less build speed impact could be found. We think of this process as a kind of ``stay-fast'' process to maintain build agility by preventing code changes negatively impacting build speed. It should raise awareness that simple changes can have significant consequences to development processes. Lebeuf et al.~\cite{Lebeuf:2018} provided a visualization framework for corresponding ``get-fast'' efforts.

In particular, we designed this approach to work in conjunction with \CloudBuild, which poses several challenges given its distributed and cached nature. In details, our approach is intended to analyze the developer's change immediately before the build, and predict: (i) whether it impacts any of the most frequent LCPs of the branch; (ii) whether the changes may lead to build time increase and an estimation of such time delta; and (iii) an estimation of the percentage of future builds that might be affected by such change and experience build time increase. The contributions of the paper can be summarized as follows:
\begin{itemize}
    \item we advocate for assisting developers in understanding the impact of their changes on build activities, so that corrective operations can be performed early in the development process;
    \item describe an approach which aims at predicting and estimating the extent to which developers' changes may impact future build activities, in terms of time and percentage of affected builds;
    \item illustrate how we plan to perform the evaluation of such predictive model.
\end{itemize}

In this paper we focus our study on Microsoft's \CloudBuild \cite{Esfahani:2016:CMD:2889160.2889222} system--a Microsoft internal cached, distributed build system. Please note that the basic concepts or \CloudBuild are very similar to those of Buck~\cite{buck} (Facebook) and Bazel~\cite{bazel} (Google). Thus, we strongly believe that the overall concepts presented in this paper are not Microsoft specific but can be applied for other build systems. However, the technical details of this work remain Microsoft specific.

\section{Problem Scale}
\label{sec:problemscale}
In this section, we provide a discussion on how painful and impacting code changes adding new dependencies can be. Please note, we are sharing Microsoft specific experiences.

In recent years, build speed regressions have become a major issue for many product teams. The \CloudBuild~team introduced a specific task-force responsible for these very expensive and time consuming investigations helping 1st party customer to overcome these issues. Some of the co-authors of this paper are part of this investigation team.

Changes to the dependency structure of the system under build represent one of the most common pattern of build time regression. In particular, adding new dependencies between existing modules or adding new modules that depend on already long dependency chains can cause build speed degradation of up to 50\%. In nearly all cases engineers were not aware of the impact their code changes would have on build speed. Removing these dependencies after the fact was nearly always painful and expensive, with cascading dependency effects.

\section{Approach}
\label{sec:approach}
In this section we describe the proposed approach, which aims at analyzing a developer's change to a code branch and estimating:

\begin{itemize}
    \item whether or not it may negatively impact future build activities;
    \item the $\Delta$ in build time increase;
    \item the percentage of future builds that might experience this time increase.
\end{itemize}
This estimation shall be performed before the actual build, when only the source code, the changes, and the Dependency Graph (DG) are available.

At first glance, it seems that such an approach could be implemented with simple checks on the LCP extracted from the DG or, even more simply, by running the build on the changed code and measuring the build time difference with the previous builds. Unfortunately, these simple implementations would not work in a distributed and cached build environment such as \CloudBuild.

\subsection{Challenges of Distributed and Cached Build Systems}
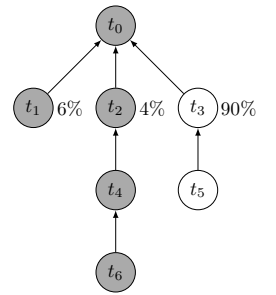
\begin{wrapfigure}{r}{0.4\linewidth}
\vspace{-0.3cm}
    \centering
    \resizebox {1.0\linewidth} {!} {
\begin{tikzpicture}[edge from parent/.style={draw,latex-},
	every node/.style={draw,circle,minimum size=1.8em, solid,black,thin},
	label distance=-4pt]
\node (t0) [fill={rgb:black,1;white,2}]{$t_0$} 
  child {node [fill={rgb:black,1;white,2}] [label=right:$6\%$] (t_1) {$t_1$}}
  child {node [fill={rgb:black,1;white,2}] [label=right:$4\%$] (t_2) {$t_2$}
    child {node [fill={rgb:black,1;white,2}] (t_4) {$t_4$}
        child {node [fill={rgb:black,1;white,2}] (t_6) {$t_6$}
  }}}
  child {node [label={right:$90\%$}] (t_3) {$t_3$}
    child {node (t_5) {$t_5$}
  }};
\end{tikzpicture}
}
\caption{Dependency Graph}
\label{fig:dg}
\vspace{-0.33cm}
\end{wrapfigure}

\figref{fig:dg} shows an example of a DG -- a directed graph representing the dependencies of a build -- where nodes represent build targets, and edges represent dependencies between targets. With \textit{target} we refer to any atomic piece of execution, such as a compilation task, a unit test, a drop \etc. When a full build is performed (\ie all the targets are executed), the unique LCP can be predicted statically by analyzing the DG and the execution time of each target. For example, in \figref{fig:dg} if we assume that each target $t_i$ has the execution time, the LCP would be the following $LCP_{full}: \{t_0, t_2, t_4, t_6\}$.

Conversely, in a cached build system such as \CloudBuild, a full build is rarely performed, since caching allows to reuse the output of a previous build and execute only a subset of the targets. For example, if the target $t_3$ is changed, the build system only needs to execute $t_3$ and all the dependent targets (\eg $t_5$), while can reuse the cached version of the remaining targets (\eg the nodes in gray in \figref{fig:dg}). In this case, the actual LCP would be $LCP: \{t_3, t_5\}$.

In such a build environment, there is not a single LCP, but rather a variety of possible LCPs depending on the changed targets and the caching status. Let $b(t_i)$ be the probability that the target $t_i$ is built, while $c(t_i) = 1 - b(t_i)$ the probability that the same target comes from cache, then we could also compute the probability of each LCP. For example, in \figref{fig:dg}, we show the probability $b(t_i)$ for the nodes $t_1, t_2, t_3$. In this example, $t_3$ is built 90\% of the times, which means that the $LCP: \{t_3, t_5\}$ is more frequent than the $LCP: \{t_2, t_4, t_6\}$. In this situation, a new dependency towards the nodes $t_3$ or $t_5$ could potentially have a greater negative effect on build activities than a new dependency on the longer $LCP_{full}$, which is rarely executed.

In summary, caching allows for optimized builds, which requires to execute only a subset of the targets. Thus, depending on which targets need to be re-built, a different LCP can be experienced for a build. The distributed environment allows multiple developers to perform builds in the cloud, which means that a developer's change could affect the build activity of many other developers.

Therefore, the impact of a software change should be measured not only on the build time increase introduced, but also on how often this time increase will be experienced, based on the probability of the affected LCP. In our preliminary analysis, we observed that the top-5 most frequent LCPs cover around 20-40\% of the total builds performed in the last 3 months for a given project. Clearly, monitoring software changes involving those LCPs would be crucial. In our proposed approach, we consider the top-$k$ most frequent LCPs, where $k$ is a project-dependent and user-defined value.

\begin{figure}[t]
\vspace{-0.0cm}
\begin{tikzpicture}[grow=right,
	level/.style={sibling distance=80mm/#1},
    level/.style ={level distance=2.5cm},
    edge from parent/.style={draw,latex-},
	every node/.style={draw,circle,minimum size=2.8em, solid,black,thin,scale=0.6},
	scale=0.6
]
\node (n1){$s_0$}
  child {node [draw=none] (n2) {$\dots$}
    	child {node (n4) {$s_j$}
        	child {node [right=0.5cm] [draw=none] (n5) {$\dots$}
            	child {node [right=0.5cm] (n6) {$s_m$} 
}}}};

\node [below=1.8cm] (t1){$t_0$}
    child {node [draw=none] (t3) {$\dots$} edge from parent[line width=1.3pt]
        child {node (t4) {$t_{i-1}$} edge from parent[line width=1.3pt]
        	child {node (t5) {$t_{i}$} edge from parent[line width=1.3pt]
        		child {node [draw=none,right=0.1cm] (t6) {$\dots$} edge from parent[line width=1.3pt]
            		child {node [right=0.1cm] (t7) {$t_n$} edge from parent[line width=1.3pt]
}}}}};

\path[draw,dashed,line width=1.3pt,-latex] (t5) -- (n4);

\end{tikzpicture}
\caption{Outward Dependency} 
\label{fig:outward}
\vspace{-0.4cm}
\end{figure}
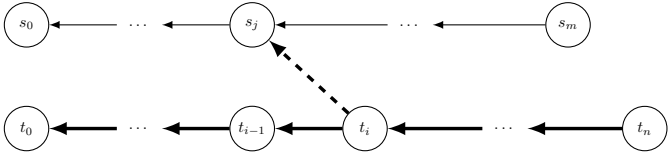

\setlength{\textfloatsep}{0pt}
\begin{algorithm}[t]
\caption{Estimation for outward dependency}
\begin{algorithmic}[1]
\Function{OutwardEstimation}{}
\If {$Time(s_0, \dots, s_j) \leq Time(t_0, \dots, t_{i-1})$}
\State $LCP \gets t_0, \dots, t_n$
\State \Return 0
\Else
\State $LCP \gets s_0, \dots, s_j, t_i, \dots, t_n$
\State \Return $\Delta Time\{ (s_0, \dots, s_j), (t_0, \dots, t_{i-1})\}$
\EndIf
\EndFunction
\end{algorithmic}
\label{alg:outward}
\end{algorithm}

\subsection{Build Impact Estimation}
Let $LCPs$ be the top-$k$ LCPs in a project, $DG_{prev}$ and $DG_{curr}$ be the DG before and after the change, the approach aims at estimating whether the change impacts one of the $LCPs$, the potential build time increase $\Delta$, and the percentage of builds affected in the future.

The approach starts by computing a graph diff between $DG_{prev}$ and $DG_{curr}$, in order to detect any new edge and node added in the change. If the diff detects a newly added edge (\ie dependency), the approach checks whether one of the two endpoints of the edge (\ie dependent or dependency node) is a target node $t_i \in LCPs$. If so, we classify the dependency in two major categories: \textit{outward} or \textit{inward} based on whether the target node $t_i$ is the dependent or dependency node.

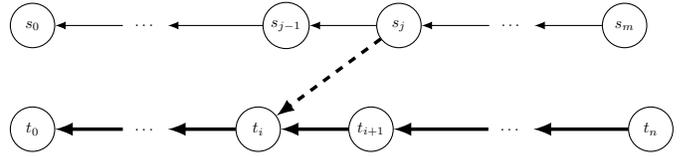
\begin{figure}[t]
\vspace{-0.0cm}
\begin{tikzpicture}[grow=right,
	level/.style={sibling distance=80mm/#1},
    level/.style ={level distance=2.5cm},
    edge from parent/.style={draw,latex-},
	every node/.style={draw,circle,minimum size=2.8em, solid,black,thin,scale=0.6},
	scale=0.6
]
\node (n1){$s_0$}
  child {node [draw=none] (n2) {$\dots$}
    child {node [right=0.1cm] (n3) {$s_{j-1}$}
    	child {node (n4) {$s_j$}
        	child {node [draw=none] (n5) {$\dots$}
            	child {node (n6) {$s_m$}
}}}}};
\node [below=1.8cm] (t1){$t_0$}
    child {node [draw=none] (t3) {$\dots$} edge from parent[line width=1.3pt]
        child {node (t4) {$t_i$} edge from parent[line width=1.3pt]
        	child {node (t5) {$t_{i+1}$} edge from parent[line width=1.3pt]
        		child {node [draw=none,right=0.1cm] (t6) {$\dots$} edge from parent[line width=1.3pt]
            		child {node [right=0.1cm] (t7) {$t_n$} edge from parent[line width=1.3pt]
}}}}};

\path[draw,dashed,line width=1.3pt,-latex] (n4) -- (t4);

\end{tikzpicture}
\caption{Inward Dependency} 
\label{fig:inward}
\vspace{-0.3cm}
\end{figure}

\setlength{\textfloatsep}{0pt}
\begin{algorithm}[t]
\caption{Estimation for inward dependency}
\begin{algorithmic}[1]
\Function{InwardEstimation}{}
\If {$Time(s_0, \dots, s_j) \geq Time(t_0, \dots, t_i)$}
\State $LCP \gets t_0, \dots, t_n$
\State \Return 0
\Else
\If {$Time(s_j, \dots, s_m) \leq Time(t_{i+1}, \dots, t_n)$}
\State $LCP \gets t_0, \dots, t_n$
\State \Return 0
\Else
\State $LCP \gets t_0, \dots, t_i, s_j, \dots, s_m$
\State \Return $\Delta Time\{ (s_j, \dots, s_m), (t_{i+1}, \dots, t_n)\}$
\EndIf
\EndIf
\EndFunction
\end{algorithmic}
\label{alg:inward}
\end{algorithm}

\subsubsection{Outward Dependency}
\figref{fig:outward} shows the $LCP: t_o, \dots, t_i, \dots, t_n$ in bold and represented as a Dependency Chain (DC), where the execution order is from left to right. The newly added outward dependency $t_i \rightarrow s_j$ is represented as a dashed edge. The node $s_j$ is also represented in its own DC, where $s_m$ is the last node to be executed in the sub-tree rooted at $s_j$ in the DG. \algoref{alg:outward} shows the steps used by the approach to estimate the potential impact of the new dependency. In algorithm, we use a proxy function $Time(\dots)$ that estimates the execution time of a sequence of build targets. If $Time(s_0, \dots, s_j) \leq Time(t_0, \dots, t_{i-1})$ (lines 2-4), the new dependency is estimated not to introduce any delay on the LCP, instead if the opposite is true (lines 5-7), the target $t_i$ will need to wait additional time before being executed (since $s_j$ has not been completed yet), therefore the new dependency is estimated to generate a new $LCP \gets s_0, \dots, s_j, t_i, \dots, t_n$. The delta in build time (line 7) will be experienced for the percentage of future builds that involve the build of the target $s_j$ (or any of the previous nodes in its DC).

Note that a special case for outward dependency is when $t_i$ is the head of the LCP. In this case, the dependency always introduce a build time delay.

\subsubsection{Inward Dependency}
\figref{fig:inward} shows the introduction of an inward dependency $s_j \rightarrow t_i$. \algoref{alg:inward} provides the pseudo-code of the steps followed by the approach in order to estimate the potential impact of the new dependency. In particular, in order for this new dependency to increase the build time, $s_j$ needs to experience a delay (\texttt{else} branch at line 5) and $Time(s_j, \dots, s_m) > Time(t_{i+1}, \dots, t_n)$ (lines 9-11). This will introduce a build time increase $\Delta$ (line 11) and generate a new $LCP \gets t_0, \dots, t_i, s_j, \dots, s_m$. The build time increase will be experienced for the future builds that involve the execution (not from cache) of target $t_i$ (or any of the previous nodes in its DC).

Note that a special case for outward dependency is when $t_i$ is the tail of the LCP. In this case, the dependency always introduce a build time delay.

\subsection{Approximation using Historical Data}
\label{sec:approximation}
The proposed approach performs its estimation by approximating execution time and probabilities using build historical data. In particular, \CloudBuild~logs execution time of each target and other metadata for every build, which can be statistically analyzed for future predictions. Execution time statistics can be used to approximate the function $Time(\dots)$, Top-$k$ most frequent LCPs can be identified by observing the build logs in the recent history of the project, similarly, probabilities $b(t_i)$ can be approximated by caching statistics.

\section{Experimental Design Plan}
\label{sec:design}
While in this paper we only formally describe the idea of the proposed approach, and we have not yet evaluated the model, this section illustrates the experimental design we intend to follow in the future. In particular, we plan to evaluate the accuracy of the approach by executing it across the change history of several software projects built using \CloudBuild, and validating its estimations using the historical build data. In details, given an historical evaluation period of a project (\eg the last 3 months of build activities), we execute the approach on each and every build submission (\ie when the build is requested, before the build is executed) and evaluate its accuracy by comparing the estimation, in the current build and the future build impact, with the real build data.

Let $b_i$ be a build in which the approach estimates that a software change generated a new $LCP_{new}$ (from the original $LCP_{old}$) which introduced a build time increase $\Delta$ and it is estimated to affect $p$\% of future builds. 

\subsection{Current Build}
In the current build session, we evaluate whether the estimated $LCP_{new}$ is actually the LCP obtained during the build execution, as reported by the historical logs and metadata. 

\subsection{Past-Future Builds}

If the current build evaluation confirms the estimation, a build time period before (\ie  \textit{past}) and after (\ie  \textit{future}) the build $b_i$ is selected approximately of the same length. Builds with $LCP_{old}$ are selected from the \textit{past} period, while builds with $LCP_{new}$ are selected from the \textit{future} period. The build execution time of the two sets is statistically analyzed in order to identify whether there is a statistically significant difference in build time, and compared it with the estimated $\Delta$. Next, the amount of builds experiencing the $LCP_{new}$ in the \textit{future} period is compared against the predicted percentage $p$\%.

\subsection{Historical Parameters}
As discussed in \ref{sec:approximation}, the approach's estimation is based on the approximations of historical data. The amount of historical data to consider is a sensible choice. On one hand, considering only few recent data points could lead to inaccuracies due to outliers, on the other hand, considering too much historical data could introduce imprecision due to data obsolescence. We plan to experiment and tune the historical parameters such as: (i) the number of most frequent LCPs $k$ and their build coverage, (ii) the historical period length when computing the targets' execution time, and (iii) the caching probabilities.

\section{Threats to Validity}
\label{sec:threats}
Threats to internal validity relate to result bias from confounding factors. The proposed approach analyzes the impact of each new dependency independently. In future work, we plan to consider the potential impact of multiple dependencies added in the same code change. Additionally, in this paper we assume that the configuration for the build environment (\ie number of machines, cores) is stable or similar across different builds.

Threats to external validity concerns the generalizability of the research. In our case, while we envisioned this system to work with \CloudBuild, the approach is generalizable to other distributed cached build systems.

\section{Conclusions}
\label{sec:conclusion}
In this paper we envision a predictive model able to alert developers on the extent to which their software changes may impact future build activities. 

As future work, we plan to evaluate the proposed approach and test its utility and usability for developers. Additionally, we plan to also incorporate positive feedback in the prediction, such as when a software change could lead to faster builds.

\section{Acknowledgment}
\label{sec:acknowledgment}
We thank K\i{}van\c{c} Mu\c{s}lu and Christian DuVarney from Tools for Software Engineers group for their valuable help.

\bibliographystyle{IEEEtran}
\bibliography{ms}

\end{document}